\documentclass[sigconf]{acmart}
\DeclareUnicodeCharacter{1EBF}{\'{e}}

\AtBeginDocument{%
  }
\renewcommand\footnotetextcopyrightpermission[1]{}

\setcopyright{acmlicensed}
\copyrightyear{2025}
\acmYear{2025}
\acmDOI{XXXXXXX.XXXXXXX}
\acmConference[ChoreoMuse]{}{Preprint}{2025}

\usepackage[capitalize]{cleveref}
\usepackage{mathtools}
\usepackage{balance}
\usepackage{multirow} 

\crefname{section}{Sec.}{Secs.}
\Crefname{section}{Section}{Sections}
\Crefname{table}{Table}{Tables}
\crefname{table}{Tab.}{Tabs.}

\begin{document}

\title{ChoreoMuse: Robust Music-to-Dance Video Generation with Style Transfer and Beat-Adherent Motion}

\author{Xuanchen Wang \enskip Heng Wang \enskip Weidong Cai \\
School of Computer Science, The University of Sydney \\
{\tt xwan0579@uni.sydney.edu.au}\\
{\tt \{heng.wang, tom.cai\}@sydney.edu.au}
}


\renewcommand{\shortauthors}{Xuanchen Wang, Heng Wang and Weidong Cai}

\begin{abstract}
Modern artistic productions increasingly demand automated choreography generation that adapts to diverse musical styles and individual dancer characteristics. Existing approaches often fail to produce high-quality dance videos that harmonize with both musical rhythm and user-defined choreography styles, limiting their applicability in real-world creative contexts. To address this gap, we introduce ChoreoMuse, a diffusion-based framework that uses SMPL format parameters and their variation version as intermediaries between music and video generation, thereby overcoming the usual constraints imposed by video resolution. Critically, ChoreoMuse supports style-controllable, high-fidelity dance video generation across diverse musical genres and individual dancer characteristics, including the flexibility to handle any reference individual at any resolution. Our method employs a novel music encoder MotionTune to capture motion cues from audio, ensuring that the generated choreography closely follows the beat and expressive qualities of the input music. To quantitatively evaluate how well the generated dances match both musical and choreographic styles, we introduce two new metrics that measure alignment with the intended stylistic cues. Extensive experiments confirm that ChoreoMuse achieves state-of-the-art performance across multiple dimensions, including video quality, beat alignment, dance diversity, and style adherence, demonstrating its potential as a robust solution for a wide range of creative applications. Video results can be found on our project page: \href{https://choreomuse.github.io}{https://choreomuse.github.io}.
\end{abstract}

\begin{teaserfigure}
  \includegraphics[width=\textwidth]{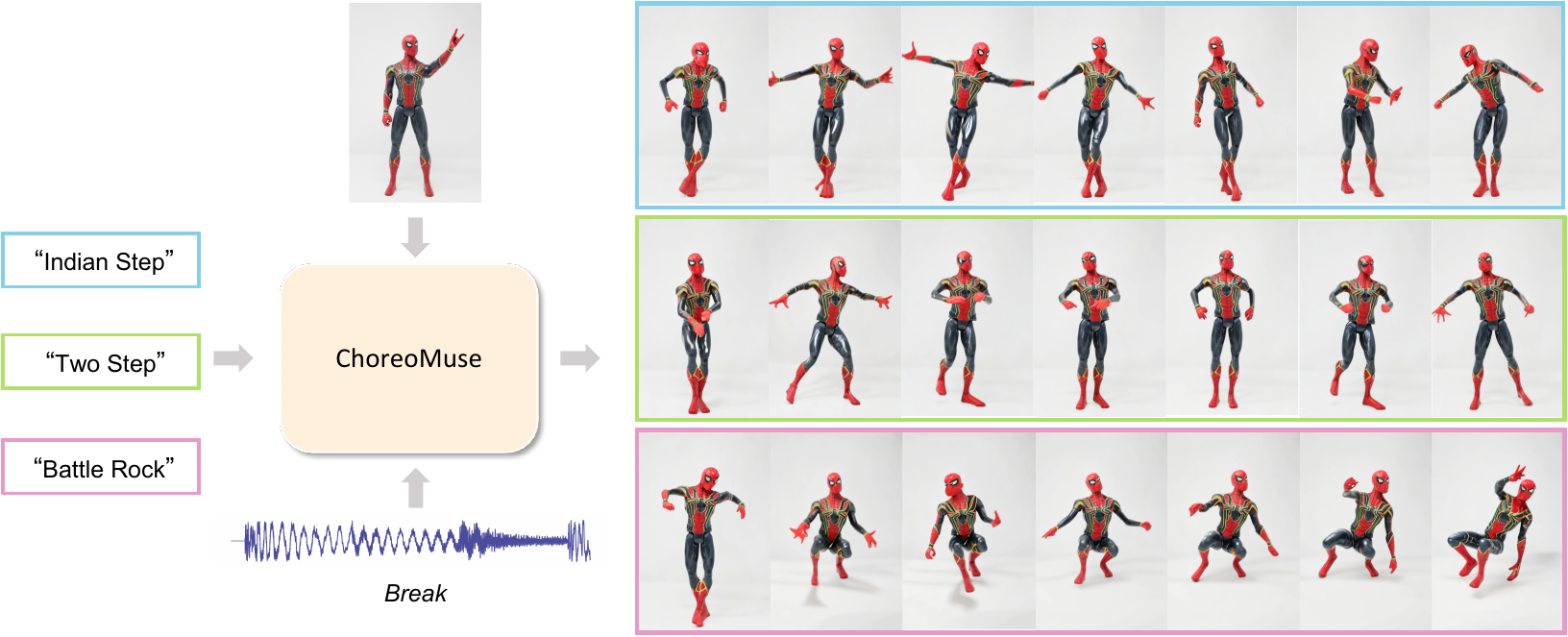}
  \caption{Examples of video frames generated by ChoreoMuse. Given a reference image and a piece of music, ChoreoMuse dynamically adjusts the choreography style to produce diverse dance videos.}
  \label{fig:starting}
\end{teaserfigure}

\keywords{Music-to-Dance Generation, Image-to-Video Generation, Diffusion Model, Multi-Modal Learning}

\maketitle

\section{Introduction}
\label{sec:intro}
Dance is a vital art form in contemporary culture, and the demand for automated choreography is rapidly expanding in creative industries such as music videos, live performances, and immersive media. Existing approaches \cite{li2021ai, siyao2022bailando, qi2023diffdance, tseng2023edge} to automated dance generation typically generate keypoint sequences, which can yield less intuitive outputs and lack support for animating specific individuals—key drawbacks for personalized or high-fidelity applications. While recent methods, such as DabFusion \cite{dabfusion}, tackle the challenge of generating dance videos from music through an optical-flow-based approach, they inevitably constrain both background variability and video resolution. Additionally, in real-world choreography, even for the same piece of music, variations in choreography style are often crucial. However, existing solutions lack the flexibility to accommodate this diversity. 

To address these shortcomings, we propose ChoreoMuse, a comprehensive, diffusion-based framework that can generate high-quality dance videos from any piece of music and reference individual. ChoreoMuse stands out for its style-controllable choreography, the flexibility to animate any reference individual in any chosen environment, and its reliance on SMPL \cite{loper2015smpl} format parameters—rather than optical flow—as the intermediate representations bridging music and video. By leveraging these parameters, ChoreoMuse overcomes resolution constraints seen in prior work, delivering sharp visuals with rich expressive details while seamlessly accommodating input reference images of any resolution and generating videos at the corresponding quality. \cref{fig:starting} showcases sample outputs: given a random reference image and a piece of music, ChoreoMuse can produce dance videos in diverse choreographic styles. ChoreoMuse’s design centers on two major training phases: 3D Dance Sequence Generation and High-Fidelity Video Generation. In the first phase, a diffusion model learns to generate 3D dance sequences in 6-DOF rotation representation \cite{zhou2019continuity}—a variant of the SMPL format—based on an audio clip and an initial pose, with a style controller offering fine-grained choreographic adjustments. In the second phase, another diffusion model creates photorealistic dance videos from a single reference image, guided by the original SMPL-based 3D dance sequence, ensuring that both the subject and the background conform to real-world aesthetic standards. Different versions of the SMPL format perform better under different conditions, so we select the optimal version for each phase. Further details on these representations are provided in \cref{sec:preliminary}. While many current methods utilize generic audio feature extractors \cite{dhariwal2020jukebox, wu2022wav2clip, wu2023large}, these pretrained models do not inherently focus on dance-relevant cues. To bridge this gap, ChoreoMuse incorporates a novel music encoder MotionTune tailored to dance generation. MotionTune is trained via a contrastive learning paradigm on paired audio and dance movement embeddings, ensuring that it aligns the temporal structure of the music with the corresponding motion patterns. 

To objectively evaluate alignment with musical style and choreographic preferences, we propose two new metrics: Music Style Alignment Score (MSAS), which measures how closely the generated dance sequences match the music’s stylistic cues, and Choreography Style Alignment Score (CSAS), which assesses whether the system’s output faithfully captures the user-defined choreographic style. These metrics offer a more comprehensive benchmark for real-life choreography scenarios. Extensive experiments demonstrate that ChoreoMuse outperforms existing methods in video quality, beat alignment, dance diversity, and style adherence. By providing a robust platform that can integrate personalization, style control, and high-quality video, ChoreoMuse holds promise for a wide range of artistic and commercial applications. Our contributions can be summarized as follows: 
\begin{itemize}
\item {}We present ChoreoMuse, a diffusion-based framework that generates high-quality dance videos in harmony with musical rhythm and user-defined choreographic styles. Extensive experiments validate that ChoreoMuse achieves state-of-the-art performance across multiple dimensions.
\item {}We present a new approach that leverages dance sequences in SMPL format to bridge music and dance videos, overcoming video resolution limitations.
\item{} ChoreoMuse employs a novel music encoder MotionTune specifically designed to capture dance-relevant cues from audio. By training MotionTune via contrastive learning on paired music and motion data, we ensure that the generated choreography closely follows the beat and expressive qualities of the input music, leading to more coherent and rhythmically aligned dance movements.
\item{} We propose Music Style Alignment Score (MSAS) and Choreography Style Alignment Score (CSAS) to objectively evaluate the fidelity of generated dances to both musical and choreographic style cues, paving the way for more robust benchmarking in automated choreography research.
\end{itemize}

\section{Related Work}
Diffusion Models (DMs)~\cite{ho2020denoising, sohl2015deep} have recently become a powerful tool in generative modeling, achieving impressive results in video generation~\cite{harvey2022flexible, ho2022imagen, ho2022video, esser2023structure, hong2022cogvideo, khachatryan2023text2video, ma2024follow}. For example, Ho et al.~\cite{ho2022video} extend the 2D U-Net into a 3D U-Net to better capture temporal dynamics, while other approaches~\cite{blattmann2023align, guo2023animatediff} build on pre-trained text-to-image models by adding temporal layers for video generation. AnimateDiff~\cite{guo2023animatediff}, for instance, introduces a motion module trained on video data that can be easily adapted to various T2I models. DMs have also been applied to image-to-video (I2V) generation. PIDM~\cite{bhunia2023person} improves pose transfer by adding texture-aware diffusion blocks, while LFDM~\cite{ni2023conditional} generates optical flow in latent space to guide image warping. DreamPose~\cite{karras2023dreampose} uses Stable Diffusion with adapters for CLIP~\cite{radford2021learning} and VAE~\cite{kingma2013auto} embeddings to enable pose-guided generation. 

At the same time, generating dance sequences from music has gained attention as a task that combines motion synthesis~\cite{aksan2019structured, butepage2017deep, hernandez2019human, holden2016deep} with music interpretation~\cite{wu2022wav2clip, mcfee2015librosa, wu2023large}, aiming to create movements that align with audio cues. Unlike traditional motion synthesis, which focuses on physical realism, dance generation requires rhythmic alignment, expressiveness, and smooth transitions. Early methods focused on 2D dance generation~\cite{shlizerman2018audio, zhuang2022music2dance, ferreira2021learning}, supported by advances in 2D pose estimation~\cite{cao2017realtime}, but lacked the depth needed for realistic choreography. This led to a shift toward 3D dance generation, using LSTMs~\cite{tang2018dance, kao2020temporally, yalta2019weakly}, GANs~\cite{lee2019dancing, sun2020deepdance, ginosar2019learning, kim2022brand}, and Transformers~\cite{huang2020dance, li2021ai, siyao2022bailando, qi2023diffdance, ye2020choreonet}. The AIST++ dataset~\cite{li2021ai} has been key in advancing this field by providing high-quality 3D dance data. Recent works include FACT~\cite{li2021ai}, which uses a full-attention transformer; Bailando~\cite{siyao2022bailando}, which combines a pose VQ-VAE with Motion GPT; and EDGE~\cite{tseng2023edge}, which supports editable dance generation. Multi-modal approaches like TM2D~\cite{gong2023tm2d} further combine music and text inputs in a shared latent space. While most studies focus on 3D motion generation, some have explored full video synthesis. Ren~\emph{et al.}~\cite{ren2020self} made an early attempt, and DabFusion~\cite{dabfusion} recently became the first to directly generate dance videos from music. However, DabFusion is limited by low resolution, static backgrounds, and poor style control. In contrast, our proposed diffusion-based framework addresses these issues, producing high-quality, style-controllable dance videos with greater flexibility and realism.

\begin{figure}[t]
  \centering
  \includegraphics[width=\linewidth]{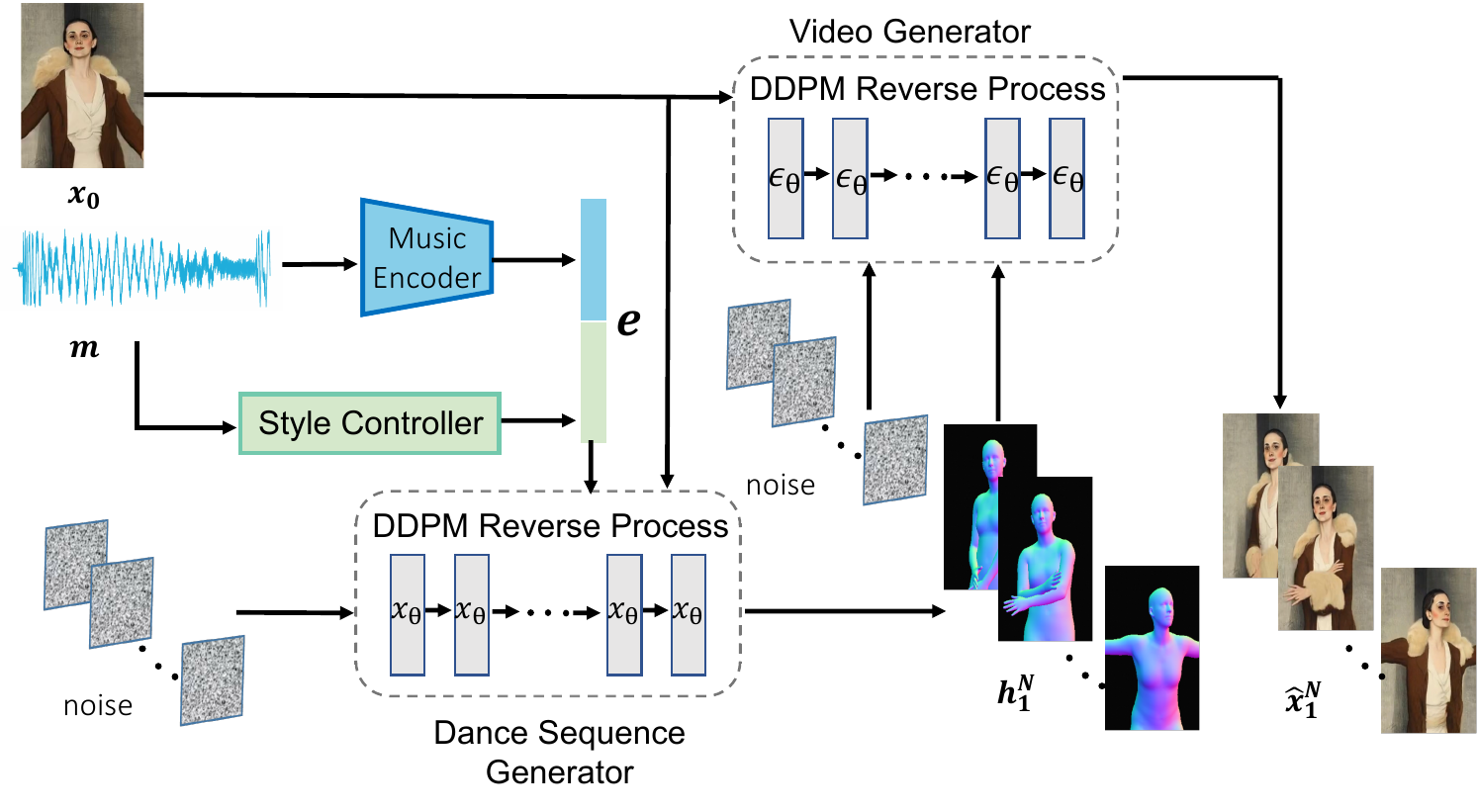}
  \caption{Inference process of ChoreoMuse. Given a piece of music $ m $ and a reference image  $ x_{0} $, the music is first processed by a music encoder to extract a music embedding and a style controller to capture choreography-related style features. These two embeddings are then fused into the final condition embedding $ \boldsymbol{e} $. The dance sequence generator produces 3D dance sequences $ h_1^N $ based on $ \boldsymbol{e} $, while the video generator synthesizes the output video $ \hat{x}_1^N $ from $ x_{0} $, guided by the generated 3D sequences.}
  \label{fig:inference_overview}
\end{figure}

\begin{figure*}[t]
  \centering
  \includegraphics[width=\textwidth]{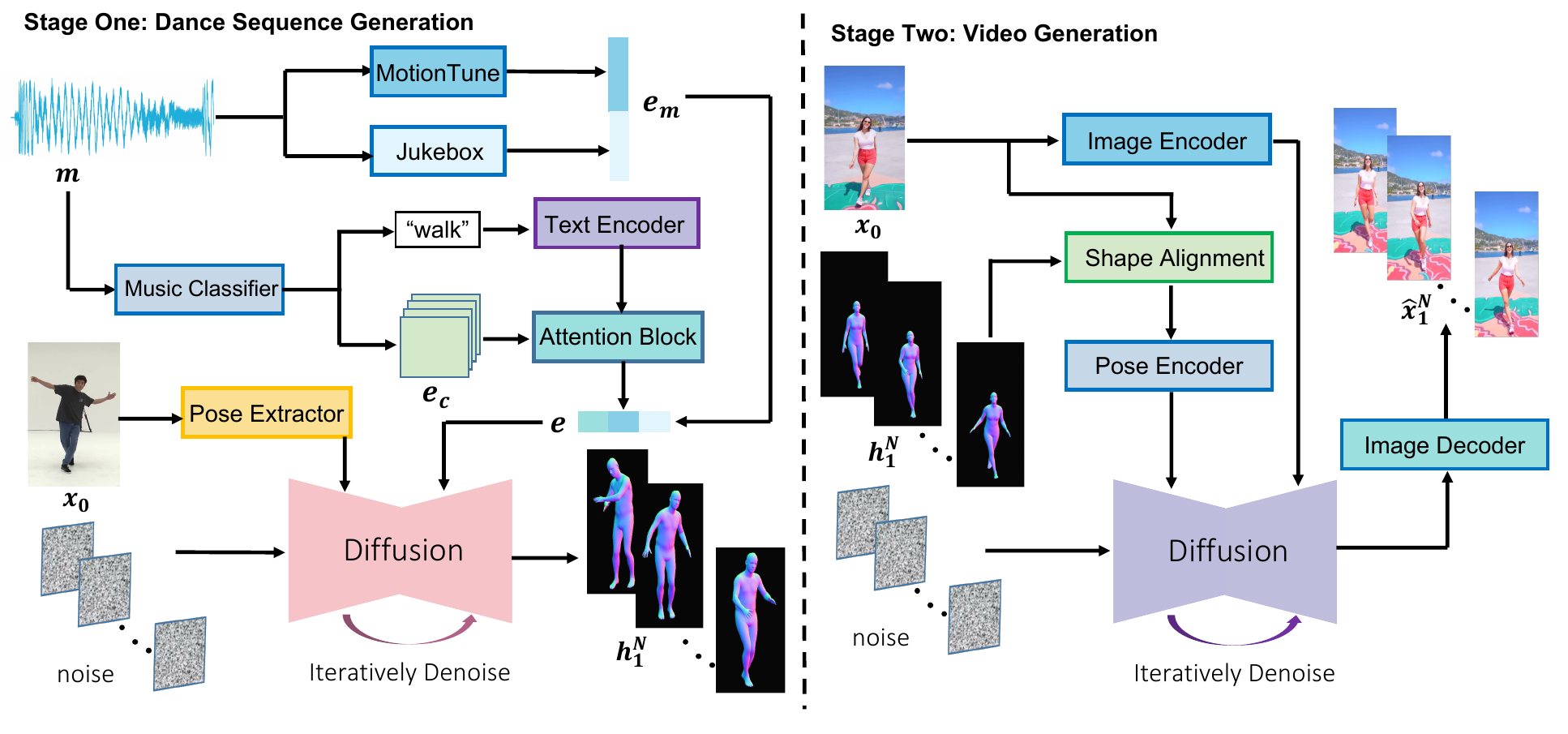}
  \caption{The training framework of ChoreoMuse. In Stage one, the dance sequence generator is trained using a reference image \(x_0\) and a music clip \(m\). Music embeddings \(\boldsymbol{e}_m\) are extracted via MotionTune and Jukebox, while a classifier predicts the music type caption to determine the choreography style. The classifier’s feature vector \(\boldsymbol{e}_c\) is combined with a text-encoded choreography style using attention blocks to generate a final style embedding, which is then merged with \(\boldsymbol{e}_m\) as input to the diffusion model. Meanwhile, a pose extractor processes \(x_0\) to create an initial pose mask for constrained training. In Stage two, the video generator is trained by first aligning \(x_0\) with the generated dance sequence $ h_1^N $. The aligned sequence is then encoded by a pose encoder and used as a condition for the diffusion model, along with an encoded reference image. Finally, the diffusion model synthesizes the complete dance video $ \hat{x}_1^N $.}
  \label{fig:training}
\end{figure*}

\section{Method}
\cref{fig:inference_overview} provides an overview of the inference pipeline in our proposed approach. Given a piece of music $ m $ and a reference image $ x_{0} $, the music first passes through a dedicated encoder that extracts a music embedding, followed by a style controller that captures choreography-related style features. These two embeddings are then combined into the final condition embedding $ \boldsymbol{e} $.  Instead of directly supplying a choreography style as input, our model relies on a classifier within the style controller to identify the music type. As shown by AIST++ \cite{li2021ai}, distinct music genres typically exhibit specialized choreography styles. For example, “POP” music often involves “hand wave” or “body wave,” whereas “House” music features choreography styles like “lose legs” or “side kick.” Once the music type is determined, the appropriate choreography style can be decided. The inference process proceeds in two key phases: dance sequence generation and video generation. In the first phase, a diffusion model employing the DDPM \cite{ho2020denoising} reverse process generates 3D dance sequences $ h_1^N $ based on $ \boldsymbol{e} $. In the second phase, another diffusion-based module synthesizes the output video $ \hat{x}_1^N $ from $ x_{0} $, guided by the generated 3D sequences. The training process of ChoreoMuse, illustrated in \cref{fig:training}, consists of two stages. A detailed explanation of each stage will be provided in the following sections. In \cref{sec:preliminary}, we introduce the latent diffusion model and detail the 3D dance sequence to establish core concepts. \cref{sec: music encoding} elaborates on the music encoding pipeline, a critical component of our framework. \cref{sec: dance generation} discusses the dance sequence generation process, and \cref{sec: video generation} explains how the final video is produced.

\subsection{Preliminary}
\label{sec:preliminary}
\noindent \textbf{Latent Diffusion Models.} We employ diffusion models \cite{ho2020denoising, sohl2015deep} across two stages to generate both 3D dance sequences and videos. Our approach follows the DDPM definition \cite{ho2020denoising} of diffusion, where a variance-preserving Markov process is applied to the latents $\{\boldsymbol{z}_t\}_{t=0}^T$. These latents represent low-dimensional features encoded from typically high-dimensional data. Specifically, in the dance sequence generation phase, they correspond to SMPL-formatted pose sequences, while in the video generation phase, they represent encoded input image. The forward noising process is given by:
\begin{equation}
    \label{eq:diffusion_forward}
    \boldsymbol{z}_t = \sqrt{\Bar{\alpha}_t}\boldsymbol{z}_0 + \sqrt{1- \Bar{\alpha}_t}\,\boldsymbol{\epsilon},
\end{equation}
where $\Bar{\alpha}_t$ is a variance schedule that determines the amount of noise added at each step, and $\boldsymbol{\epsilon} \sim \mathcal{N}(0,\mathcal{I})$. We set $T=1000$ timesteps. The reverse diffusion process then reconstructs the original data from noise. We defer the specifics of the denoising process and the loss function to later sections, where we cover each training stage individually.

\noindent \textbf{3D Dance Sequence.} To bridge music and video, we adopt 3D dance sequences represented in SMPL \cite{loper2015smpl} format. A primary motivation for selecting SMPL-format dance sequences as the link between the two stages is the rich, structured information they provide, which effectively guides dance video generation. SMPL is a widely used parametric 3D human body model in computer vision, graphics, and animation, designed to generate realistically deforming human body shapes from a relatively small set of parameters. It uses low-dimensional pose parameters $\theta \in \mathbb{R}^{24 \times 3 \times 3}$ and shape parameters $\beta \in \mathbb{R}^{10}$ to produce a 3D mesh $M \in \mathbb{R}^{3 \times N}$ with $N=6890$ vertices. A vertex-wise weight matrix $W \in \mathbb{R}^{N \times k}$ encodes the relationship between each vertex and the body joints $J \in \mathbb{R}^{3 \times k}$, enabling human part segmentation. In the original SMPL model, each joint rotation is parameterized by an axis-angle representation (three parameters per joint), which can introduce discontinuities and singularities. More recent approaches \cite{guo2022generating, petrovich2021action, petrovich2022temos} adopt a 6-DOF (6D) rotation representation to avoid issues like wrap-around and gimbal lock. According to our experiments, the original SMPL format is compact and straightforward to interpret, so we use it in the video generation stage where high-quality pose data largely mitigates discontinuity-related problems. For the dance sequence generation stage, however, avoiding artifacts such as foot-skating is crucial; hence, we use the 6D rotation representation for every joint plus a single root translation, resulting in $w \in \mathbb{R}^{24 \cdot 6 + 3 = 147}$. We further add binary contact labels for the heel and toe of each foot to control foot–ground interactions. Thus, the total pose representation is $x \in \mathbb{R}^{4 + 147 = 151}$.

\subsection{Music Encoding}
\label{sec: music encoding}
Extracting meaningful information from music is crucial in our approach. Existing models \cite{tseng2023edge, dabfusion, qi2023diffdance} typically rely on a pretrained large-scale network to extract music features, capturing only general information. However, our task demands more nuanced mappings between music and dance movements to ensure higher-quality choreography. Consequently, in addition to using a pretrained large model, we train a specialized music encoder MotionTune. 

We employ two encoders to separately process the music data $X_i^m$ and the pose data $X_i^p$, where $(X_i^m, X_i^p)$ is one of the music-pose pairs indexed by $i$. We also incorporate a textual input $X_i^t$, formed by combining the music genre and the choreography style (e.g., ``House: walk out''). The music embedding $E_i^m$ is extracted by the audio encoder $f_{\text{music}}(\cdot)$ and projected via:
\begin{equation}
    E_i^{m} = \mathrm{MLP}_{\text{music}} \bigl(f_{\text{music}}(X_i^m)\bigr).
\end{equation}
Meanwhile, the dance embedding $E_i^d$ is obtained by projecting the concatenated outputs of the pose encoder $f_{\text{pose}}(\cdot)$ and the text encoder $f_{\text{text}}(\cdot)$:
\begin{equation}
    E_i^{d} = \mathrm{MLP}_{\text{dance}}
    \Bigl( 
      \bigl[ 
        f_{\text{pose}}(X_i^p),\,
        f_{\text{text}}(X_i^t) 
      \bigr] 
    \Bigr).
\end{equation}

We train the model using a contrastive learning framework \cite{elizalde2023clap} that aligns music and dance embeddings in pairs, following:
\begin{align}
\begin{split}
\resizebox{\columnwidth}{!}{
    $\mathcal{L} = \frac{1}{2N} \sum_{i=1}^N \biggl(\log \frac{\exp(E_i^m \cdot E_i^d / \tau)}{\sum_{j=1}^N \exp(E_i^m \cdot E_j^d / \tau)}
    + \log \frac{\exp(E_i^d \cdot E_i^m / \tau)}{\sum_{j=1}^N \exp(E_i^d \cdot E_j^m / \tau)}\biggr),$
}
\end{split}
\end{align}
where $\tau$ is a learnable temperature parameter, and $N$ denotes the batch size during training. Additionally, we incorporate output features from Jukebox \cite{dhariwal2020jukebox}---a large GPT-style model trained on one million songs to generate raw music audio---because it provides strong, general-purpose audio representations. As illustrated in \cref{fig:training} (left), we fuse these features with those from MotionTune to form the final music embeddings $ \boldsymbol{e_m} $.

\subsection{Dance Sequence Generation}
\label{sec: dance generation}
\cref{fig:training} (left) shows an overview of the training process for this stage. In this stage, our objective is to generate 3D dance sequences $ \hat{x}_1^N $ from the given music $ m $ and an initial pose extracted from the reference image $ x_0 $. To guide dance sequence generation, music embeddings \(\boldsymbol{e}_m\) are first extracted using MotionTune and Jukebox. A classifier then predicts the music genre caption, producing a feature vector \(\boldsymbol{e}_c\), which is fused with a text-encoded choreography style via attention blocks to generate a style embedding. This style embedding is combined with \(\boldsymbol{e}_m\) and fed into the diffusion model. Meanwhile, a pose extractor processes \(x_0\) to produce an initial pose mask, serving as a constraint during training.

\noindent \textbf{Loss Function.}
Following the diffusion approach described in \cref{sec:preliminary}, we apply noise to the pose data \(x\) and then reverse the forward diffusion process. Specifically, we train a denoising neural network \(x_\theta(\boldsymbol{z}_t, t, c)\) (with model parameters \(\theta\)) to recover \(x\) at each time step \(t\). Here, \(c\) serves as the condition, combining both music embeddings and choreography style embeddings. Our basic loss function is:
\begin{equation} \label{eq:2}
\mathcal{L}_\text{basic} = \mathbb{E}_{x, t}\Bigl[\| x - x_\theta(\boldsymbol{z}_t, t, c)\|_2^2\Bigr].
\end{equation}

Beyond this basic loss, we adopt auxiliary losses, following \cite{tseng2023edge}, to improve physical realism. These auxiliary losses encourage the alignment of three aspects of physical plausibility: joint positions, velocities, and foot velocities:
\begin{equation}
\mathcal{L}_\text{joint} = \frac{1}{N} \sum_{i=1}^{N} \bigl\| FK\bigl(x^{(i)}\bigr) - FK\bigl(x_\theta^{(i)}\bigr)\bigr\|_{2}^{2},
\end{equation}
\begin{equation}
\mathcal{L}_\text{vel} = \frac{1}{N-1} \sum_{i=1}^{N-1} \bigl\| \bigl(x^{(i+1)} - x^{(i)}\bigr) - \bigl(x_\theta^{(i+1)} - x_\theta^{(i)}\bigr)\bigr\|_{2}^{2},
\end{equation}
\begin{equation}
\mathcal{L}_\text{foot} = \frac{1}{N-1} \sum_{i=1}^{N-1} \bigl\| \bigl(FK\bigl(x_\theta^{(i+1)}\bigr) - FK\bigl(x_\theta^{(i)}\bigr)\bigr) \cdot \hat{b}^{(i)}\bigr\|_{2}^{2},
\end{equation}
where \(FK(\cdot)\) denotes the forward kinematic function, converting joint angles into joint positions, \(x^{(i)}\) indicates the \(i\)-th frame, and \(\hat{b}^{(i)}\) is the model’s predicted binary foot-contact label at frame \(i\).

The overall training loss is a weighted sum of the basic objective and the auxiliary losses:
\begin{equation} \label{eq:6}
\mathcal{L_\text{AC}} = \mathcal{L}_\text{basic} \;+\; \lambda_\text{pos}\,\mathcal{L}_\text{joint} \;+\; \lambda_\text{vel}\,\mathcal{L}_\text{vel} \;+\; \lambda_\text{foot}\,\mathcal{L}_\text{foot}.
\end{equation}

\noindent \textbf{Choreography Style Controller.}
In real-world scenarios, the length of a music piece can vary. We often split the music into smaller segments for generation. However, choreography style consistency is critical: even if we slice the input audio, the choreography style should remain the same throughout. Consequently, the Choreography Style Controller (CSC) must handle variable-length audio inputs. We first train a music classifier to recognize the music style, producing a style caption and feature vectors (from its last two layers) that enhance the model’s ability to differentiate choreography styles. 

To manage variable-length inputs within a constant computational budget, we adopt a Feature Fusion Mechanism \cite{wu2023large}. For an audio clip of length \(T\) seconds and a fixed chunk duration \(d=5\) seconds, if \(T \leq d\), we repeat the input and then pad it with zeros. If \(T > d\), we compress the input to \(d\) seconds for a global view, then randomly slice three \(d\)-second clips from the front, middle, and back one-third of the original input for local details. These four segments (\(4 \times d\)) are fed into the initial audio encoder layer to obtain features. The three local features are merged via a 2D convolution layer (stride 3 in the time dimension), and the resulting local feature \(F_{\text{local}}\) is fused with the global feature \(F_{\text{global}}\) as:
\begin{equation}
F_{\text{fusion}} = \alpha \,F_{\text{global}} + (1-\alpha)\,F_{\text{local}},
\end{equation}
where \(\alpha = f_{\mathrm{AFF}}\bigl(F_{\text{global}}, F_{\text{local}}\bigr)\) is computed by an attention-based feature fusion module \cite{dai2021attentional}. The classifier outputs both a style caption (used to determine choreography style) and combined features \(\boldsymbol{e}_c\). We then encode the textual choreography style to obtain an embedding \(\boldsymbol{e}_t\). We select two models, CLIP transformer \cite{radford2021learning} and BERT \cite{devlin2019bert} to construct the text encoder. Finally, \(\boldsymbol{e}_t\) passes through an MLP layer to interact with \(\boldsymbol{e}_c\) via self- and cross-attention layers, producing the final choreography style embedding. This embedding is then combined with the music embedding to serve as the conditioning input $ \boldsymbol{e} $ for the diffusion model.

\noindent \textbf{Initial Pose Constraint.}
When a reference image is provided, we must consider its corresponding initial pose in the final choreography. Rather than encoding the initial pose as an additional condition—which risks introducing interference—we instead adopt a standard masked denoising approach \cite{lugmayr2022repaint}. Given the initial pose \(x^{\text{start}}\) and a binary mask \(m\) specifying known regions, at each denoising step we apply:
\begin{equation} \label{eq:8}
x_{t-1} \coloneqq m \odot q\bigl(x^{\text{start}}, t-1\bigr) \;+\; (1 - m) \odot x_{t-1},
\end{equation}
where \(q(\cdot)\) is the forward noising process from Eq.~(\ref{eq:diffusion_forward}), and \(\odot\) denotes element-wise multiplication. This replaces the known pose regions with forward-diffused samples consistent with the initial pose constraint. This approach also allows for generating dance sequences of arbitrary length. For instance, although our model is trained on 5-second clips, we can generate a 7.5-second clip by constraining the first 2.5 seconds of the new sequence to match the last 2.5 seconds of the previous clip.

\noindent \textbf{Diffusion Architecture.}
The diffusion architecture for this stage begins with a linear projection layer, followed by four attention blocks. Each attention block contains one self-attention layer, two feature-wise linear modulation (FiLM)~\cite{perez2018film} layers, a cross-attention layer, two additional FiLM layers, an MLP layer, and a final FiLM layer. A final projection layer is applied after these four attention blocks.

\subsection{Video Generation}
\label{sec: video generation}
\cref{fig:training} (right) illustrates an overview of this stage’s training. The video generator is trained by first aligning the initial pose \(x_0\) with the generated dance sequence \(h_1^N\) from stage one. This aligned sequence is then passed through a pose encoder to produce a conditioning representation for the diffusion model. Along with an encoded reference image, this representation is used to guide the diffusion model in synthesizing the complete dance video \(\hat{x}_1^N\).


\noindent \textbf{Loss Function.}
Our objective in this stage is to generate dance videos conditioned on both a reference image and the guidance of the previously generated 3D dance sequence. Unlike the first stage, here the denoising process predicts noise \(\boldsymbol{\epsilon}_{\theta}(\boldsymbol{z}_t,t,\boldsymbol{c})\) for each timestep \(t\) when transitioning from \(\boldsymbol{z}_t\) to \(\boldsymbol{z}_{t-1}\). The function \(\boldsymbol{\epsilon}_{\theta}\) denotes a noise prediction network, with the dance sequence serving as the condition \(\boldsymbol{c}\). The loss function measures the expected mean squared error (MSE) between the true noise \(\boldsymbol{\epsilon}\) and the predicted noise \(\boldsymbol{\epsilon}_{\theta}\):
\begin{equation}
\mathcal{L}_\text{VG} \;=\; \mathbb{E}_{\boldsymbol{\epsilon},\,t}
\bigl[\|\boldsymbol{\epsilon} \;-\; \boldsymbol{\epsilon}_\theta(\boldsymbol{z}_t,\,t,\,\boldsymbol{c})\|_2^2\bigr].
\end{equation}

\noindent \textbf{Shape Alignment.}
A critical challenge in human video generation is animating a reference image according to a motion sequence while preserving the subject’s original shape and appearance. To address this, we propose a silhouette-based shape alignment strategy, refining the SMPL body parameters by ensuring consistency between the rendered and observed silhouettes. Let \(\beta\) and \(\theta\) represent the SMPL shape and pose parameters, respectively. Given a reference image \(I_{\text{ref}}\), we obtain a foreground mask \(S_{\text{ref}}(\mathbf{u}) \in \{0,1\}\) at each pixel \(\mathbf{u}\) via a person-segmentation network. We then render the SMPL mesh to produce a binary silhouette \(S_{\text{SMPL}}(\beta,\theta)(\mathbf{u})\), where the silhouette is 1 if \(\mathbf{u}\) projects inside the mesh, and 0 otherwise. We define a pixel-wise silhouette loss:
\begin{equation}
\label{eq:silhouette_loss}
\mathcal{L}_{\mathrm{sil}}(\beta, \theta)
\;=\;
\sum_{\mathbf{u}}
\Bigl\|
  S_{\text{SMPL}}(\beta,\theta)(\mathbf{u})
  \;-\;
  S_{\text{ref}}(\mathbf{u})
\Bigr\|^2.
\end{equation}
This term penalizes discrepancies between the SMPL-rendered silhouette and the ground-truth foreground mask. In practice, we combine it with a keypoint alignment term:
\begin{equation}
\label{eq:resized_alignment}
\resizebox{0.9\hsize}{!}{$
\displaystyle
\min_{\beta,\theta}
\biggl[
 \lambda_{\mathrm{kpt}}
 \sum_{i}
 \bigl\|
   p_i
   \;-\;
   \pi\bigl(\beta,\theta, i\bigr)
 \bigr\|^2
 \;+\;
 \lambda_{\mathrm{sil}}\,\mathcal{L}_{\mathrm{sil}}\bigl(\beta,\theta\bigr)
 \;
\biggr],
$}
\end{equation}
where \(p_i\) denotes the \(i\)-th 2D keypoint in the reference image, \(\pi(\beta,\theta,i)\) is a projection function giving the \(i\)-th keypoint’s 2D location, and \(\{\lambda_{\mathrm{kpt}}, \lambda_{\mathrm{sil}}\}\) are weighting factors. By jointly optimizing these shape \(\beta\) and pose \(\theta\) parameters, we achieve better silhouette alignment, capturing the reference subject’s unique body contours where skeletal constraints alone are insufficient.

\noindent \textbf{Dance Sequence Encoding.}
To guide video generation effectively, we render the transferred SMPL mesh into multiple 2D representations, including depth maps (for camera-to-pixel distance), normal maps (for surface orientations), semantic segmentation maps (for pixel-wise body part labels), and foot contact masks (converted from binary foot-contact labels). We then fuse these rendered features using Multi-Layer Motion Fusion~\cite{zhu2024champ}, producing a single, unified guidance signal for video generation.

\noindent \textbf{Diffusion Architecture.}
In this stage of diffusion, we employ a U-Net-based model following a standard encoder-decoder structure with skip connections. The encoder downsamples the noisy latent representation using convolutional layers, incorporating motion guidance through cross-attention. A self-attention block at the bottleneck fuses motion features and ensures temporal consistency across frames. The decoder then upsamples the features, leveraging skip connections to restore details. Additionally, a temporal attention module smooths motion transitions, while ReferenceNet helps maintain identity consistency. 




\begin{table*}[t]
\centering
\begin{tabular}{c|cccc|c}
    \toprule
    Method & PSNR $\uparrow$ & SSIM $\uparrow$ & LPIPS $\downarrow$ & FVD $\downarrow$ & Dataset \\
    \midrule
    \multicolumn{6}{c}{\textit{Video Generation from Music}} \\
    \midrule
    DabFusion~\cite{dabfusion} & 25.98  &  0.895  & 0.325 & 195.2 & \multirow{2}{*}{AIST++} \\
    Ours                        & \textbf{27.85} & \textbf{0.934} & \textbf{0.243} & \textbf{178.6} & \\
    \midrule
    \multicolumn{6}{c}{\textit{Video Generation from Poses}} \\
    \midrule
    DisCo~\cite{wang2024disco}         & 28.85  &  0.659  & 0.298 & 296.5 & \multirow{4}{*}{TikTok} \\
    MagicAnimate~\cite{xu2024magicanimate} & 29.25  &  0.729  & 0.237 & 176.3 & \\
    Animate Anyone~\cite{hu2024animate}    & 29.49  &  0.715  & 0.291 & 173.5 & \\
    Ours                                 & \textbf{29.85} & \textbf{0.755} & \textbf{0.233} & \textbf{165.4} & \\
    \bottomrule
\end{tabular}

\caption{Quantitative video quality comparisons on the AIST++ and TikTok datasets.}
\label{tab:tiktok comparison}
\end{table*}

\begin{table*}[t]
\centering
 \begin{tabular}{c|cccccc}
    \toprule
    Method          & PFC $\downarrow$ & $ \text{Dist}_k $ $\uparrow$ & $ \text{Dist}_g $ $\uparrow$ & BAS $\uparrow$ &MSAS $\uparrow$&CSAS $\uparrow$ \\ \midrule
    \textit{Bailando}~\cite{siyao2022bailando}  & 1.759  &  7.85  & 7.69 & 0.23& 0.69&-\\
    FACT~\cite{li2021ai}  & 2.193  &  10.83  & 6.09 & 0.22&0.72&-\\
    DiffDance~\cite{qi2023diffdance}  & 1.698  &  6.21  & 3.02 & 0.24&0.73&-\\
    EDGE~\cite{tseng2023edge}  & \textbf{1.542}  &  9.53 & 5.75 & 0.26&0.75&-\\
    DabFusion~\cite{dabfusion}  & 2.231  &  8.52  & 6.59 & 0.23&0.75&-\\
    Ours  & \textbf{1.542}          & \textbf{11.08}           & \textbf{7.92}          & \textbf{0.28} & \textbf{0.81}& \textbf{0.84} \\ \bottomrule
\end{tabular} 
\caption{Quantitative dance quality and style alignment comparisons on the AIST++ test set.}
  \label{tab:dance comparison}
\end{table*}

\begin{table}[t]
\centering
\resizebox{\linewidth}{!}{

 \begin{tabular}{c|cccccc}
    \toprule
    Method          & PFC $\downarrow$ & $ \text{Dist}_k $ $\uparrow$ & $ \text{Dist}_g $ $\uparrow$ & BAS $\uparrow$ &MSAS $\uparrow$&CSAS $\uparrow$ \\ \midrule
    w/o. MotionTune  & 1.576  &  9.89  & 6.85 & 0.23&0.76&0.81\\
    Ours  & \textbf{1.542}          & \textbf{11.08}           & \textbf{7.92}          & \textbf{0.28} & \textbf{0.81}& \textbf{0.84} \\ \bottomrule
\end{tabular} 
}
\caption{Ablation study on the impact of MotionTune. “w/o. MotionTune” indicates the absence of MotionTune.}
  \label{tab:motiontune comparison}
\end{table}

\begin{table}[t]
\centering

  \resizebox{\linewidth}{!}{
 \begin{tabular}{c|cccccc}
    \toprule
    Method          & PFC $\downarrow$ & $ \text{Dist}_k $ $\uparrow$ & $ \text{Dist}_g $ $\uparrow$ & BAS $\uparrow$ &MSAS $\uparrow$&CSAS $\uparrow$ \\ \midrule
    Original SMPL  & 1.587  &  10.03  & 7.25 & 0.26&0.80&0.83\\
    6D Rotation  & \textbf{1.545}          & \textbf{11.08}           & \textbf{7.92}          & \textbf{0.28} & \textbf{0.81}& \textbf{0.84} \\ \bottomrule
\end{tabular} 
}
\caption{Ablation study comparing 6D rotation and original SMPL in dance sequence generation.}
  \label{tab:smpl d}
\end{table}

\begin{table}[t]

  \resizebox{\linewidth}{!}{
 \begin{tabular}{c|cccc}
    \toprule
    Method          & PSNR $\uparrow$ & SSIM $\uparrow$ & LPIPS $\downarrow$ & FVD $\downarrow$ \\ \midrule
    6D Rotation  & 26.87  &  0.904  & 0.288 & 180.2\\
    Original SMPL  & \textbf{27.85}          & \textbf{0.934}           & \textbf{0.243}          & \textbf{178.6}\\ \bottomrule
\end{tabular}

}
\caption{Ablation study comparing 6D rotation and original SMPL in video generation.}
\label{tab:smpl v}
\end{table}

\section{Experiments}
\subsection{Dataset and Implementation Details}
We use the AIST++ \cite{li2021ai} dataset, which is curated by professional dancers and includes expert-designed choreography, to train both MotionTune and the dance sequence generation stage. AIST++ comprises 1,408 high-quality dance motions paired with music spanning diverse genres. For training the video generator, however, explicit music labels and choreography are not required; rather, the emphasis is on motion quality. Consequently, we employ the dataset from \cite{zhu2024champ}, which contains approximately 5,000 high-fidelity human videos—sourced from reputable online repositories—totaling about 1 million frames. We additionally use the TikTok dataset~\cite{jafarian2021learning} for comparison with other models. This dataset contains over 100,000 RGB images extracted at 30 FPS from more than 300 manually curated TikTok dance videos, each featuring a single person performing dances. It also includes segmentation and UV coordinate annotations, enabling high-fidelity human depth learning.

Our model is trained using 4 NVIDIA A100 GPUs. In MotionTune, we select PANN \cite{kong2020panns} and HTSAT \cite{chen2022hts} as the core audio encoders.  For the pose encoder, we use the model from MotionCLIP~\cite{tevet2022motionclip}, which is a transformer auto-encoder with eight layers. All audio files are preprocessed to a mono channel at a 48kHz sampling rate in FLAC format. In the dance sequence generation stage of ChoreoMuse, training examples are clipped to 5-second segments (30 FPS). In the video generation stage, we sample, resize, and center-crop individual video frames to a uniform resolution of \(768\times768\). During inference, we employ a temporal aggregation method to achieve continuity in longer sequences, allowing smooth integration of results across distinct batches. Additional training details are provided in the supplementary materials.

\begin{figure}[t]
  \centering
  \includegraphics[width=\linewidth]{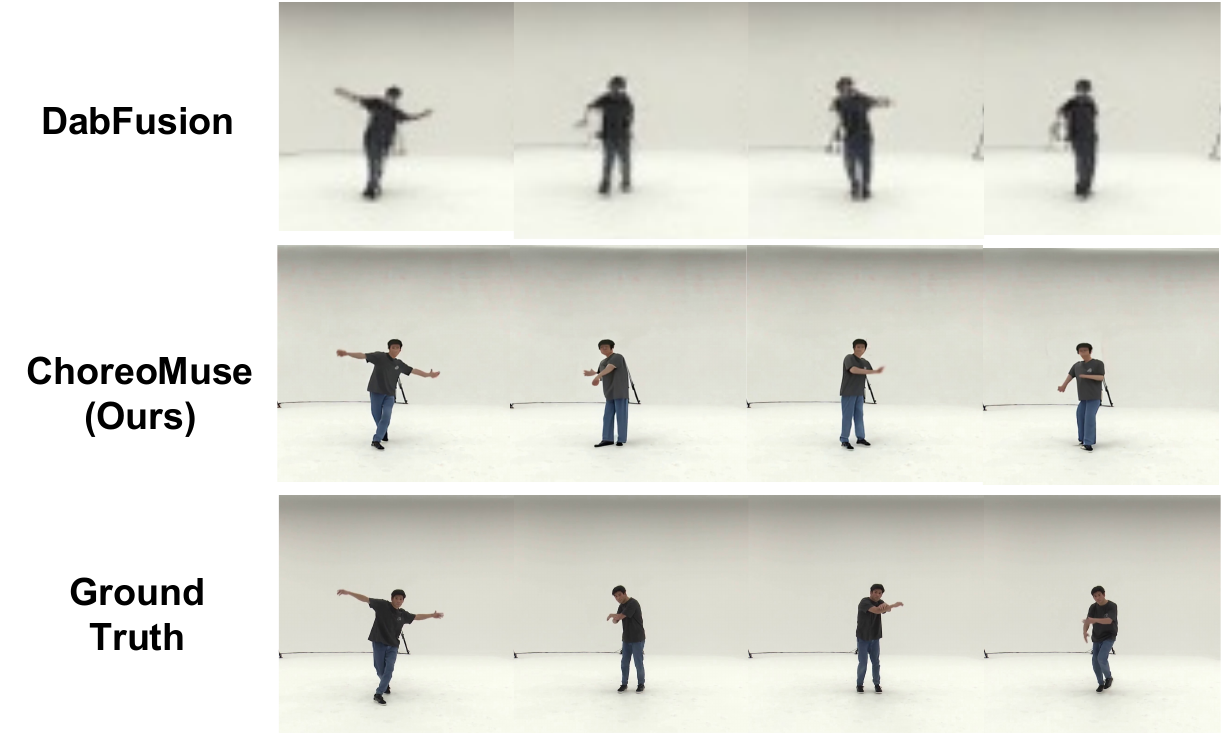}
  \caption{Qualitative comparisons between ChoreoMuse and DabFusion on the AIST++ test set.}
  \label{fig:comparison}
\end{figure}

\subsection{Baselines}
Music-to-dance video generation is a relatively new research area; to our knowledge, DabFusion~\cite{dabfusion} is the only existing model that directly tackles this task. To quantitatively demonstrate the superior quality of our model’s outputs, we also compare our method to several state-of-the-art image animation approaches that, while not designed specifically for music-to-dance generation, can help illustrate the quality of our results. In particular, we select DisCo~\cite{wang2024disco}, MagicAnimate~\cite{xu2024magicanimate}, and Animate Anyone~\cite{hu2024animate} as additional baselines. For fairness, we compare against DabFusion on the AIST++ test set, and against DisCo, MagicAnimate, and Animate Anyone on the TikTok dataset~\cite{jafarian2021learning}. For choreographic quality, we compare not only to DabFusion but also to three state-of-the-art motion generation models—FACT~\cite{li2021ai}, Bailando~\cite{siyao2022bailando}, DiffDance~\cite{qi2023diffdance}, and EDGE~\cite{tseng2023edge}—to more thoroughly evaluate how well our method captures dance movements and styles.

\subsection{Evaluation Metrics}

\noindent \textbf{Video Quality.}
We assess both single-frame quality and overall video fidelity. For single-frame quality, we measure Structural Similarity Index (SSIM)~\cite{wang2004image}, Learned Perceptual Image Patch Similarity (LPIPS)~\cite{zhang2018unreasonable}, and Peak Signal-to-Noise Ratio (PSNR)~\cite{hore2010image}. Video fidelity is evaluated using Fréchet Video Distance (FVD)~\cite{unterthiner2018towards}.

\noindent \textbf{Choreography Quality.}
We evaluate choreography quality along three dimensions. First, we measure physical plausibility using the Physical Foot Contact (PFC) score \cite{tseng2023edge}. Second, we quantify diversity in generated dances by calculating distributional spreads in both the \emph{kinetic} (Dist$_k$) and \emph{geometric} (Dist$_g$) feature spaces as in \cite{tseng2023edge}. Lastly, we gauge music-motion alignment with Beat Alignment Scores (BAS) \cite{siyao2022bailando}.

\noindent \textbf{Music Style Alignment Score (MSAS).}
To evaluate how well a generated dance sequence aligns with a music style, we train a multi-class style classifier \(f\). Given a dance sequence \(x_i\), this classifier outputs probabilities \(P(s \mid x_i)\) over a set of music styles \(\mathcal{S}\). For each test pair \((x_i, s_i)\), where \(s_i \in \mathcal{S}\) is the ground-truth style, we consider the top-3 most likely styles predicted by \(f\bigl(x_i\bigr)\). Denote these as \(\mathrm{Top3}\bigl(x_i\bigr)\). If the true style \(s_i\) is among the top-3, we take \(P\bigl(s_i \mid x_i\bigr)\); otherwise, the score is zero:
\begin{equation}
\text{Score}_{\mathrm{top3}}(x_i, s_i)
~=\;
\begin{cases}
P\bigl(s_i \mid x_i\bigr), & \text{if } s_i \in \mathrm{Top3}\bigl(x_i\bigr),\\
0, & \text{otherwise}.
\end{cases}
\end{equation}
The final MSAS metric is the average of these scores over all \(N\) test examples:
\begin{equation}
    \text{MSAS}
~=\;
\frac{1}{N}\,\sum_{i=1}^{N}
\text{Score}_{\mathrm{top3}}(x_i, s_i).
\end{equation}
We adopt a top-3 criterion rather than a strict top-1 classification because real-world dances often exhibit overlapping styles, making style classification inherently ambiguous.

\noindent \textbf{Choreography Style Alignment Score (CSAS).}
In many cases, each broad music style can encompass multiple choreographic sub-styles, and collecting sufficient labeled data for a large multi-class classifier is impractical. Instead, we propose a distance-based approach to measure alignment without requiring an extensive supervised dataset. Define \(\phi(x)\in\mathbb{R}^d\) as the feature embedding for a dance sequence \(x\). For a choreography style \(c\), suppose we have a reference set of real dances \(\{x_1^{(c)}, x_2^{(c)}, \ldots\}\), forming \(\mathcal{D}_c\). We compute the centroid:
\begin{equation}
\mu_c
=\;
\frac{1}{|\mathcal{D}_c|}
\sum_{x_j^{(c)}\in\,\mathcal{D}_c}
\phi\bigl(x_j^{(c)}\bigr).
\end{equation}
Given a test set of \(N\) generated dances \(\{(x_i,c_i)\}_{i=1}^N\), where \(c_i\) is the intended choreographic style, we define the style alignment score as:
\begin{equation}
\text{CSAS}
~=\;
\frac{1}{N}\,
\sum_{i=1}^N
\exp\!\Bigl(-\,\alpha\,\bigl\|\phi(x_i) \;-\; \mu_{c_i}\bigr\|\Bigr),
\end{equation}
where \(\|\phi(x)-\mu_c\|\) is the Euclidean distance indicating how closely \(x\) matches style \(c\), and \(\alpha>0\) is a decay parameter. Lower distances yield higher alignment scores, ensuring generated dances better reflect the intended sub-style without relying on a fully supervised model for every class.

\begin{figure*}[t]
  \centering
  \includegraphics[width=\textwidth]{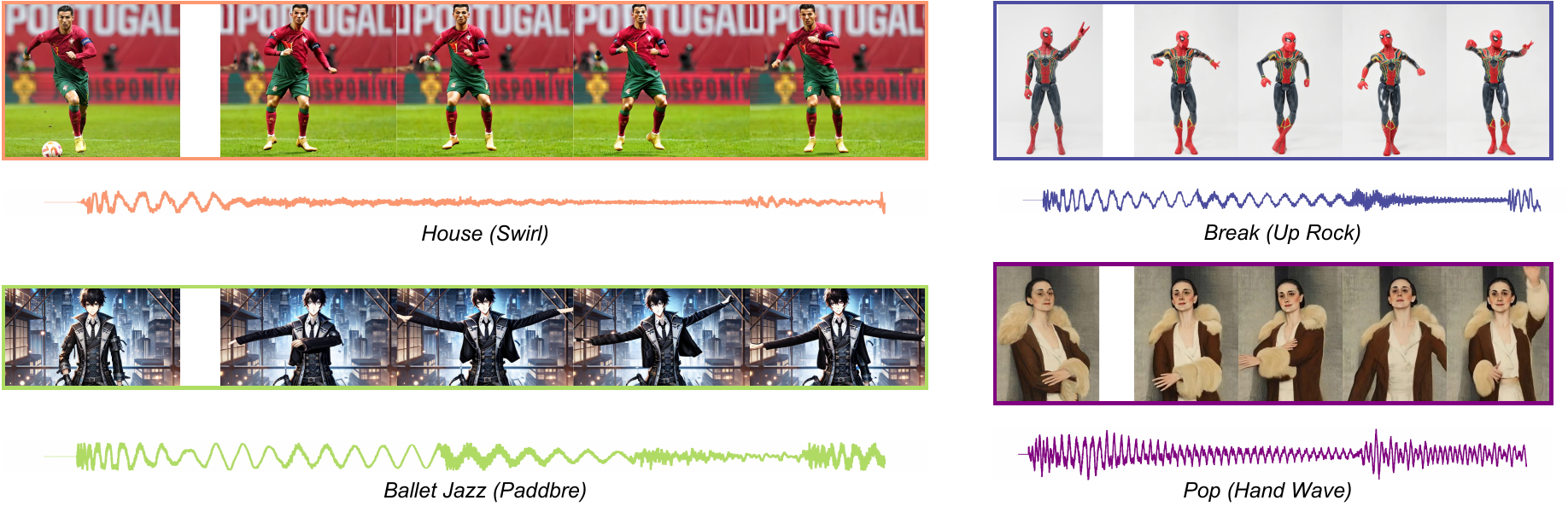}
  \caption{Generated video frames from ChoreoMuse, demonstrating its ability to process input images of any resolution and accommodate any subject. }
  \label{fig:showcase}
\end{figure*}

\subsection{Result Analysis}
\noindent \textbf{Video Quality.}
\cref{tab:tiktok comparison} provides a quantitative comparison of DabFusion and ChoreoMuse on the AIST++ dataset, showing that ChoreoMuse outperforms DabFusion on all reported metrics. In addition, it compares ChoreoMuse to cutting-edge image-animation methods on the TikTok dataset, where our approach again achieves higher performance on every metric. \cref{fig:comparison} illustrates qualitative results from the AIST++ test set, revealing that our generated videos more closely resemble ground truth than those produced by DabFusion.

\noindent \textbf{Choreography Quality and Style Alignment.}
\cref{tab:dance comparison} summarizes the results for choreography quality and style alignment across various methods. ChoreoMuse achieves the highest scores for PFC, Dist\textsubscript{k}, Dist\textsubscript{g}, MSAS, and BAS. Notably, the other approaches lack user-controlled choreography style, so we propose our CSAS metric as a baseline for future style-controllable research.

\noindent \textbf{Versatility of ChoreoMuse.}
\cref{fig:showcase} demonstrates the broad applicability of ChoreoMuse: it accepts reference images at different resolutions and outputs same-resolution videos. The model also supports a wide variety of subject types, ranging from real humans to toys, comic characters, and even oil-painting figures, without sacrificing video quality.

\noindent \textbf{Human Evaluations.}
Since ChoreoMuse supports both style-controllable generation and newly introduced style alignment metrics, we conducted a user study to validate these metrics. We recruited 125 participants: 20 choreographers, 40 dancers, and 65 individuals without formal dance or choreography experience. Each participant viewed 100 video clips generated by our model, paired with corresponding ground-truth videos, to assess alignment with both music style and choreography style. Specifically, there were 5 ground-truth videos referencing the same music style and 3 referencing the same choreography style. If a participant judged that at least 80\% of the videos matched a particular style, we counted this as a successful alignment. According to the results, 104 participants (83.2\%) found the music-style match successful, while 96 participants (76.8\%) affirmed the choreography-style match. These findings underscore the strong alignment capabilities of ChoreoMuse in both music and choreography style dimensions. Additional user study details are provided in the supplementary materials.

\subsection{Ablation Study}
\noindent \textbf{Effectiveness of MotionTune.}
We conduct an ablation study to assess the influence of our MotionTune encoder by comparing model performance with and without MotionTune embeddings (see \cref{tab:motiontune comparison}). Our results show that incorporating MotionTune boosts the model’s performance across all dance-quality metrics, indicating that its specialized music-to-dance representation significantly enhances choreography generation.

\noindent \textbf{Impact of SMPL Parameter Formats.}
We further investigate the role of different SMPL parameter formats in the two training stages. For dance sequence generation, \cref{tab:smpl d} demonstrates that the 6D rotation representation is more effective. This suggests that tailoring the SMPL parameter format to each specific stage yields optimal performance. For video generation, \cref{tab:smpl v} shows that the original SMPL format delivers superior results, presumably due to its compactness and straightforward mapping to visual features. 




\section{Conclusion}
In this paper, we introduce ChoreoMuse, a novel diffusion-based framework for generating high-fidelity dance videos conditioned on both music and a reference image. Unlike previous approaches, ChoreoMuse not only captures musical rhythms and motion patterns but also offers fine-grained choreography style control. To achieve this, we leverage SMPL parameters as an intermediate 3D representation of human pose and shape. Our method follows a two-stage training paradigm. It first generates a dance sequence from the input music, then synthesizes a corresponding video guided by the reference image. We also propose MotionTune, a new music encoder that extracts dance-relevant information from audio. To quantify how well generated dances align with both musical and choreographic styles, we introduce two new evaluation metrics. Extensive experiments demonstrate that ChoreoMuse achieves state-of-the-art results across multiple benchmarks.


\bibliographystyle{ACM-Reference-Format}
\balance
\bibliography{main}

\appendix

\end{document}